# Behavior of the anomalous correlation function in uniform 2D Bose gas

Abdelâali Boudjemâa[1]

*Department of Physics, Faculty of Sciences, Hassiba Benbouali University of Chlef*
*P.O. Box 151, 02000, Chlef, Algeria.*

**Abstract**

We investigate the behavior of the anomalous correlation function in two dimensional Bose gas. In the local case, we find that this quantity has a finite value in the limit of weak interactions at zero temperature. The effects of the anomalous density on some thermodynamic quantities are also considered. These effects can modify in particular the chemical potential, the ground sate energy, the depletion and the superfluid fraction. Our predictions are in good agreement with recent analytical and numerical calculations. We show also that the anomalous density presents a significant importance compared to the non-condensed one at zero temperature. The single-particle anomalous correlation function is expressed in two dimensional homogenous Bose gases by using the density-phase fluctuation. We then confirm that the anomalous average accompanies in analogous manner the true condensate at zero temperature while it does not exist at finite temperature.

---

[1] Corresponding author :e-mail: **a.boudjemaa@univ-chlef.dz**

## I. Introduction

The experimental progress of the ultracold gases in two dimensions (2d) [1-8] has recently attracted great attention. The properties of these fluids are radically different from those in three dimensions. The famous Mermin-Wagner-Hohenberg theorem [9, 10] states that long-wavelength thermal fluctuations destroy long-range order in a homogeneous one-dimensional Bose gas at all temperatures and in a homogeneous two-dimensional Bose gas at any nonzero temperature, preventing formation of condensate.

Since the earlier works of Schick [11] and Popov [12], several theoretical studies of fluctuations, scattering properties and the appropriate thermodynamics have been performed in [13-16]. In fact, in most of the previous references, the anomalous density is neglected under the claim that it is a divergent and unmeasured quantity as well as its contribution is very small compared to the other terms. Otherwise, the importance of the anomalous density in three-dimensional Bose gas has been shown in our recent theoretical results [17, 18] and also by several authors [19-24] using different approaches. Theoretically, the anomalous average arises of the symmetry-breaking assumption [17, 19, 24]. It quantifies the correlations of pairs of non-condensate atoms with pairs of condensate atoms due to the Bogoliubov pair promotion process in which two condensate atoms scatter each other out of the condensate which is responsible for the well-known Bogoliubov particle-hole structure of excitations in the system [24]. The anomalous density can also be interpreted as a measure of the squeezing of the non-condensate field fluctuations [25]. Certainly, the presence of this quantity adds new features to the well-known problems and attracts our attention to the two-dimensional systems. A number of questions arise naturally in this paper: Does the anomalous density exist even at finite temperature in 2d Bose gas? How does its behavior compare with the normal density at zero temperature? What are the effects of this quantity on the thermodynamics of the system?

Due to the complexity and the particularity of dilute 2d Bose gases, many analytical investigations have been performed recently to find corrections beyond mean-field at zero temperature. One should cite at this stage that Pricoupenko [26] employs the pseudo potential with a Gaussian variational approach. Mora and Castin [27], on the other hand, used their lattice model which is a sort of regularization scheme to treat ultraviolet divergences. Cherny *et al* [28] used a reduced-density matrix of second order and a variational procedure to derive results identical to those of Refs [26, 27] for equation of state (EoS) and ground-state energy. The above analytical results have been checked using Monte Carlo calculations to find numerical agreement with beyond mean-field terms in 2d [29, 30]. Recently, Mora and Castin [31] have been also extended their approach [27] one step beyond Bogoliubov theory which gives good accuracy with the simulations of [30]. Another kind of extensions has been developed recently by Sinner *et al* [32] which based on using the functional renormalization group to study dynamical properties of the 2d Bose gas at $T=0$. The approach is free from infrared divergences and satisfies both the Hungeholtz-Pines (HP) [33] relation and the Nepomnyashchy identity [34], which states that the anomalous self-energy vanishes at zero frequency and momentum. The spectrum energy thus satisfies a Bogoliubov-type expression with a renormalized sound velocity. Although the above approaches provide good predictions for the thermodynamic of 2d Bose gas in the universal regime, they are limited only at zero temperature.

The present paper deals with extending our variational Time Dependent Hartree-Fock-Bogoliubov (TDHFB) theory to the case of 2d Bose systems. The theory was previously presented for 3d systems in [17, 18]. In fact, the main difference between our approach and the earlier variational HFB treatments is that in our variational theory we do not minimize only the expectation values of a single operator like the free energy in the standard HFB approximation. Conversely, our variational theory is based on the minimization of an action also with a Gaussian variational ansatz. The action to minimize involves two types of variational objects: one related to the observables of interest and the other akin to a density matrix [35, 36].

The paper is organized as follows: In Sec. II, we briefly review the derivation of the TDHFB formalism, and give the different quantities which we study in 2d homogeneous system. In Sec. III, we restrict ourselves to the behavior of the anomalous density and its effects on the depletion, the chemical potential and the ground-state energy. We therefore, compare our results with recent Monte-Carlo simulations and analytic predictions. The validity of the HP theorem and Nepomnyashchy identity are also discussed within the present formalism. In Sec. IV, we extend our results at finite temperature where we calculate in particular the one-body anomalous correlation function. In Sec. V, we apply our formalism to analyze the behavior of the superfluid fraction. We then emphasize the importance of the anomalous density for the occurrence of superfluid transition and sound velocity. Our conclusion and perspectives are drawn in Sec. VI.

## II. Formalism

Our starting point is the TDHFB equations which describe the dynamics of $d$-dimensional interacting trapped Bose systems. For a short-range interaction potential and sufficiently dilute gas, the TDHFB equations read.

$$i\hbar \dot{\Phi}(r) = \left[ h^{sp} + gn_c(r) + 2g_d \tilde{n}(r) \right] \Phi(r) + g_d \tilde{m}(r) \Phi^*(r), \qquad (1.a)$$

$$i\hbar \frac{d\rho}{dt} = \Im \rho - \rho \Im^+, \qquad (1.b)$$

where $h^{sp} = -\frac{\hbar^2}{2m}\Delta + V_{\text{ext}}(\vec{r}) - \mu$ is the single-particle Hamiltonian, $V_{\text{ext}}(\vec{r})$ is the external trapping potential, $\mu$ is the chemical potential and $g_d$ is the interaction parameter in $d$-dimensions.

Here we have defined the 2×2 matrices

$$\Im(r.r') = \begin{pmatrix} h(r,r') & \Delta(r',r') \\ -\Delta^*(r,r) & -h^*(r,r') \end{pmatrix}$$

and

$$\rho(r.r') = \begin{pmatrix} \tilde{n}(r,r') & \tilde{m}(r,r') \\ \tilde{m}^*(r,r') & \tilde{n}^*(r,r') + \delta(r,r') \end{pmatrix},$$

where $h(r,r') = h^{sp}(r) + 2g_d \left[ \tilde{n}(\vec{r}',\vec{r}) + \Phi^*(\vec{r}')\Phi(\vec{r}') \right]$, $\Delta(r,r) = g_d \left[ \tilde{m}(\vec{r},\vec{r}) + \Phi(\vec{r})\Phi(\vec{r}) \right]$ and

$$\begin{aligned} \tilde{n}(\vec{r},\vec{r}') &\equiv \tilde{n}^*(\vec{r},\vec{r}') = <\Psi^+(\vec{r})\Psi(\vec{r}')> - \Phi^*(\vec{r})\Phi(\vec{r}') \\ \tilde{m}(\vec{r},\vec{r}') &\equiv \tilde{m}(\vec{r}',\vec{r}) = <\Psi(\vec{r})\Psi(\vec{r}')> - \Phi(\vec{r})\Phi(\vec{r}') \end{aligned}, \qquad (2)$$

are respectively the normal and anomalous single-particle correlation functions. In the local case they play the role of the non-condensed and anomalous densities. Moreover, our formalism provides a direct link between these two later quantities as

$$I(\vec{r},\vec{r}') = \int d\vec{r}'' \left[ \rho_{11}(\vec{r},\vec{r}'')\rho_{22}(\vec{r}'',\vec{r}') - \rho_{12}(\vec{r},\vec{r}'')\rho_{21}(\vec{r}'',\vec{r}') \right]. \qquad (3)$$

Notice that Eq.(3) is often known as the Heisenberg invariant, it is a direct consequence of the conservation of the von-Neumann entropy $S = -\text{Tr}\,\mathcal{D}\ln\mathcal{D}$. For pure state Eq.(3) takes the form $I(\vec{r},\vec{r}') = \delta^d(\vec{r} - \vec{r}')$ [37].

Among the advantages of the TDHFB equations is that the three densities are coupled in a consistent and closed way. Second, they should in principle yield the general time, space and temperature dependence of the various densities. Furthermore, they satisfy the energy and number conserving laws. In addition, the most important feature of the TDHFB equations is that they are valid for any Hamiltonian $H$ and for any density matrix operator. Interestingly, our TDHFB equations can be extended to provide self-consistent equations of motion for the triplet correlation function by using the post Gaussian ansatz.

In the uniform case ($V_{ext}(r) = 0$) and for a thermal distribution at equilibrium, by working in the momentum space,

$$\rho_{ij}(\vec{r}, \vec{r}') = \int \frac{d^d k}{(2\pi)^d} e^{i\vec{k}.(\vec{r}-\vec{r}')} \rho_{ij}(k), \qquad (4)$$

where $\rho_{ij}(k)$ is the Fourier transform of $\rho_{ij}(\vec{r}, \vec{r}')$. We can then easily rewrite Eq. (3) as

$$I_k = \tilde{n}_k(\tilde{n}_k + 1) - |\tilde{m}_k|^2 = \frac{1}{4\sinh^2(\varepsilon_k / 2T)}, \qquad (5)$$

with $\varepsilon_k$ is the Bogoliubov energy spectrum given below.

The physical meaning of Eq.(5) is that it allows us to calculate in a very useful way the dissipated heat for $d$-dimensional Bose gas as

$$Q_d = \frac{1}{n} \int E_k I_k \frac{d^d k}{(2\pi)^d}, \qquad (6)$$

where $E_k = \hbar^2 k^2 / 2m$ is the energy of a free particle.

Furthermore, at zero temperature, Eq. (5) reduces to $|\tilde{m}_k|^2 = \tilde{n}_k(\tilde{n}_k + 1)$, which constitutes an explicit relationship between the normal and the anomalous densities at zero temperature and indicates that these two quantities are of the same order of magnitude at low temperatures which leads to the fact that neglecting $\tilde{m}$ while maintaining $\tilde{n}$ is a quite unsafe approximation.

The excitation energy $\varepsilon_k$ is determined in our formalism via the random-phase approximation (RPA) [36] which can be found by expanding all quantities around their equilibrium solution. The RPA appears as a direct application of the general Balian-Vénéroni formalism to the Lie algebra of single boson operators [36, 37]. Thus we write

$$\Phi(k,t) = \Phi(k) + \delta\Phi(k,t)$$
$$\tilde{n}(k,t) = \tilde{n}(k) + \delta\tilde{n}(k,t) \quad , \tag{7}$$
$$\tilde{m}(k,t) = \tilde{m}(k) + \delta\tilde{m}(k,t)$$

Then, we have written these quantities on a diagonal basis (RPA matrix) which derived from the set (3), and kept only the first-order terms. After a long, but straightforward calculation, we arrive at the gapless expression of the Bogoliubov spectrum [18, 38]

$$\varepsilon_k = \sqrt{(E_k - \mu + \Sigma_{11})^2 - \Sigma_{12}^2}, \tag{8}$$

where $\Sigma_{11} = 2g_d n$ and $\Sigma_{12} = g_d(n_c + \tilde{m})$ are respectively the first order normal and anomalous self-energies with $n = n_c + \tilde{n}$ is the total density.

A detailed derivation of the Bogoliubov spectrum with the RPA method will be given elsewhere.

Note that Eq.(8) can be also obtained using the Green's functions (see e.g.[38]). It provides a useful finite-temperature version of the healing length and the sound velocity $c_s$ as

$$\xi = \hbar/\sqrt{m\Sigma_{12}} = \hbar/\sqrt{mn_c g_d \left(1 + \frac{\tilde{m}}{n_c}\right)} = \hbar/mc_s, \tag{9}$$

In order to get explicit formulas for the non-condensed and the anomalous averages in $d$-dimensions we may use Eq. (5). A simple calculation yields

$$\tilde{n} = \frac{1}{2}\int \frac{d^d k}{(2\pi L^{-1})^d}\left[\frac{E_k + g_d(n_c + \tilde{m})}{\varepsilon_k}\sqrt{I_k} - 1\right], \tag{10.a}$$

$$\tilde{m} = -\frac{1}{2}\int \frac{d^d k}{(2\pi L^{-1})^d}\left[\frac{g_d(n_c + \tilde{m})}{\varepsilon_k}\sqrt{I_k}\right], \tag{10.b}$$

It is worth noting that Eqs. (1.a) and (10) together form the generalized HFB equations. This shows that, in the static case, our formalism recovers easily the full HFB equations at both finite and zero temperatures.

### III. Anomalous density at zero temperature

Let us now discuss the behavior of the normal and anomalous densities in homogeneous 2d Bose gas at both zero and finite temperatures. From this point we consider the regime in weakly repulsive interaction at zero temperature where $\sqrt{I_k} = 1$. In two-dimensional Bose gas, the interaction parameter ($g_d = g_2$) depends logarithmically on the chemical potential as

$$g_2 = \left[\frac{4\pi\hbar^2}{m}\frac{1}{\ln(2\hbar^2/m\mu a^2)}\right], \tag{11}$$

where $a$ is the two-dimensional scattering length among the particles and $g_2$ is the two-body $T$-matrix (see e.g. [15,26,27]).

The calculation of the integral in Eq. (10.a) leads us to the following expression of the depletion

$$\frac{\tilde{n}}{n} = \frac{1}{4\pi n \xi^2}. \tag{12}$$

This equation is in good agreement with that obtained in [26].

The integral in Eq.(10.b) has an ultraviolet divergence in both two and three dimensions. This divergence is well-known and arises due to the use of the contact potential. To regulate the ultraviolet divergences, we may use the dimensional regularization [39-40]. In such a technique one calculates the loop integrals in $d = 2 - 2\eta$ dimensions for values of $\eta$ where the integrals converge. One then analytically continues back to $d = 2$ dimensions. With dimensional regularization, an arbitrary renormalization scale $M$ is introduced. This scale can be identified with the simple momentum cutoff. An advantage of dimensional regularization is that in two dimensions systems it automatically sets power divergences to zero, while logarithmic divergences show up as poles in $\eta$ [40]. Using this technique one gets for the anomalous density

$$\tilde{m}_{T=0}^{\Lambda} = -\frac{\Lambda^2}{4} J_{0,1}, \tag{13}$$

where $\Lambda$ is the regularized-part which is related to the size of particles and interactions as $\Lambda = 2/\xi$. This parameter is similar to that used in [26, 32, 40].

And $J_{0,1} = \frac{1}{4\pi}\left[\frac{1}{\eta} - L + O(\eta)\right]$ with $L = \ln(\Lambda^2/4M^2)$.

Thus the convergent part of the anomalous density provides

$$\tilde{m}_{T=0}^{\Lambda} = \frac{\Lambda^2}{16\pi}\ln(\Lambda^2/4M^2). \tag{14}$$

A useful remark at this level, the non-condensed density of Eq.(12) can be rewritten also in terms of $\Lambda$ as $\tilde{n} = \Lambda^2/16\pi$.

At $T = 0$, the condensed density has a significant value and hence constitutes the dominant quantity in the system, while both the non-condensed and the anomalous densities vanish for $\Lambda \to 0$ which ensures that $\Sigma_{11}^{\Lambda=0} = 2\mu$ and $\Sigma_{12}^{\Lambda=0} = \mu$ in good accordance with the Hugenholtz-Pines theorem $\Sigma_{11}^{\Lambda=0} - \Sigma_{12}^{\Lambda=0} = \mu$ [33]. On the other hand, once we find $\Sigma_{12}^{\Lambda=0} \neq 0$, this means that the actual version of our extended variational TDHFB including a dimensional regularization does not satisfy the Nepomnyashchy identity [32] as it should be for any limited approximation order (see for example Griffin and Shi [38], Yukalov [19], Pricoupenko [26] and Andersen [40]). Indeed, in a Bose-condensed system, the anomalous self-energy must be nonzero in order to define a meaningful nonzero sound velocity and healing length. In addition, a zero sound velocity leads evidently to an unstable system (see Eq.(9)). Indeed, our approach can be an effective way to verify the Nepomnyashchy identity, but on the condition that we sum over all terms of perturbation theory for the self-energy with renormalization of the sound velocity to ensure the stabilization of the system as it has been demonstrated in [32].

Conversely, at finite-temperature three-dimensional Bose gas, the chemical potential satisfies the generalized version of HP theorem given by Hohenberg and Martin [41] $\Sigma_{11}^{\Lambda} - \Sigma_{12}^{\Lambda} = \mu$. In such a situation, when $T \geq T_c$, both the condensate and the anomalous densities vanish [17-19, 24, 25] whatever the value of $\Lambda$, which implies directly that $\Sigma_{12}^{\Lambda} = 0$ and $\Sigma_{11}^{\Lambda} = 2g\tilde{n} \approx 2\mu$. Consequently, a vanishing anomalous self-energy is further guaranteed at high temperature and momentum in 3d systems. Physically, this result is reasonable because the gas becomes completely thermalized and therefore there is neither superfluid nor acoustic waves when the temperature reaches its critical value.

On the other side, the dimensional regularization gives an asymptotically exact result at weak interactions ($g_2 \to 0$). The extrapolation to finite interactions requires that the limiting condition $\tilde{m}/n_c \ll 1$ be verified. In real systems, however, the interactions have a finite range $a$ and so $1/a$ provides a natural ultraviolet cutoff $M$ [40]. Hence, using this technique presupposes that value of the healing length in Eqs.(12) and (13) takes the form $\xi = \hbar / \sqrt{mn_c g_2}$. In the case where $n_c \approx n$, we recover easily the well-known result of Schick [11] for the depletion, while Eq.(14) has no analog in the literature. Under these conditions, the anomalous average turns out to be given as

$$\tilde{m}_{T=0} = \frac{m\mu_0}{4\pi\hbar^2} \ln\left(\frac{m\mu_0 a^2}{\hbar^2}\right), \tag{15}$$

where $\mu_0 = g_{2d} n$.

FIG.1 shows that the non-condensed and the anomalous densities, as function of the dimensionless parameter $x = m\mu_0 a^2 / \hbar^2$, are competitive contributions at zero temperature. For small values of $x$, we observe that the non-condensed density is greater than the anomalous one while this later becomes the dominant quantity for the whole range of $x$ ($x > 10$). $\tilde{n}$ and $\tilde{m}$ are comparable only for $x \approx 10$. Therefore, we deduce that omitting the anomalous density, while keeping the normal one is physically and mathematically inappropriate. It is noticed that this behavior holds also in three dimensional Bose gas [17-25].

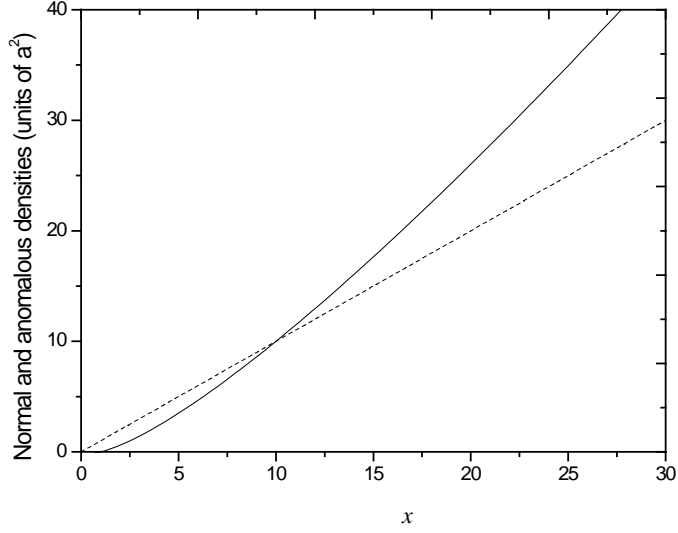

**FIG 1. Non-condensed (dashed line) and anomalous (solid line) densities as function of $x$**

It is important now to discuss how the anomalous density can modify the chemical potential and hence the other thermodynamic quantities of dilute Bose gas. The first order quantum corrections to the chemical potential are given by $\delta\mu = g_d (\tilde{n} + \tilde{m})$ [42]. Therefore, using the results obtained in Eqs.(12) and (15) with the assumption $n \approx n_c$ at $T = 0$. One obtains for the chemical potential

$$\mu = \mu_0 \left[ 1 + \frac{m\mu_0}{4\pi\hbar^2 n} \ln\left( \frac{m\mu_0 a^2 e}{\hbar^2} \right) \right], \quad (16)$$

The ground-state energy is obtained through

$$E/N = \int_0^n \mu \, dn = \frac{\mu_0}{2} \left[ 1 + \frac{mg_2}{8\pi\hbar^2} \ln\left( \frac{m\mu_0 a^2 \sqrt{e}}{\hbar^2} \right) \right], \quad (17)$$

The leading term in Eq.(16) was first obtained by Schick [11] while the second represents our correction to the chemical potential. Clearly this correction is *universal*, depending only on the interactions and scattering length. It is worth mentioning that the additional logarithm term in Eq.(16) is analogous to that found recently in [31]. Moreover, what is interesting in Eq. (16) is that if we invert it and take the limit of vanishing density, we thereby recover the well-known Popov's EoS [12].

Before plotting figures (2) and (3), we use the dimensionless relation $na^2 = x^2 \left[ \frac{1}{2} - \ln\left( \frac{xe^\gamma}{2} \right) \right]$ where $\gamma$ is the Euler's constant [26, 31].

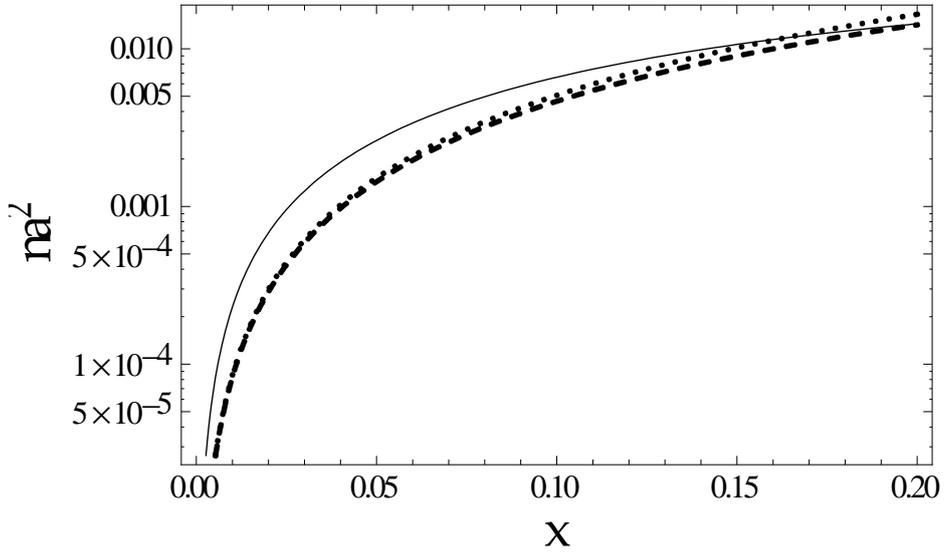

**FIG. 2. Equation of State of 2D homogeneous Bose gas. Solid line: our extended variational approach. Dashed line: EoS predicted in [26, 27]. Dotted line: first correction beyond Bogoliubov theory [31].**

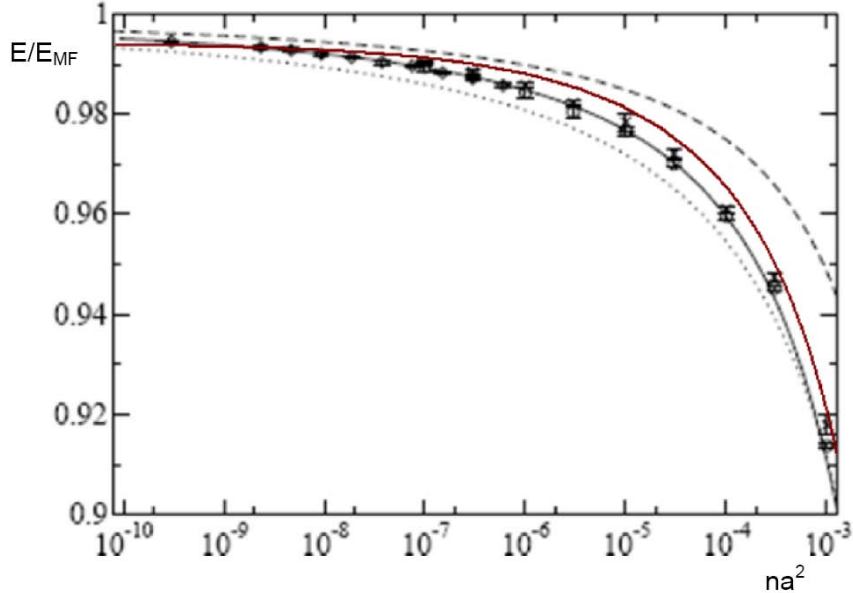

FIG. 3. Ground state energy of a two dimensional Bose gas, as a function of the gas density, in units of the mean field prediction $E_{MF}$. Brown solid line: our calculations. Dark solid line: energy obtained from the beyond Bogoliubov [31]. Dashed line: analytical prediction of [28]. Dotted line: analytical calculations of [26]. Plotting symbols with error bars: numerical results of [29], for interactions given by hard disks (crosses) and by soft disks (circles); numerical results of [30], for dipolar interactions (diamonds).

One can see from FIG.2 that for the value of the gas parameter around $5\times 10^{-5}$ there is a difference of 10% between the equations of state while for $na^2$ larger than $5\times 10^{-3}$ all EoS are practically identical.

FIG.3 shows that our expression of the ground sate energy (Eq.(17)) gives also an estimate compatibility within error bars of Monte-Carlo simulations [29, 30] and analytic results of [26-28, 31]. Furthermore, it is clearly seen from FIGs.2 and 3 that there is an upward shift of our curve relative to that of ref [31], which is indeed due to a prefactor which appears in the expansion of the later reference.

Another important feature revealed in FIG.3 is that there is no difference between dipolar and short-range interaction for densities lower than $10^{-10}$ [30].

## IV. Anomalous density at finite temperature

Now we turn to analyze the finite temperature case. As we mentioned in the introduction, the finite temperature uniform 2d Bose gas is characterized by the absence of a true Bose-Einstein condensate and long-range order [9, 10]. So the physics of 2d Bose gas at finite-$T$ can be understood in the context of the density-phase representation. Accordingly, the single-particle anomalous correlation function is found by using the field operator in the form: $\Psi = \sqrt{n} e^{i\phi}$. Following the hydrodynamic approach described in [14, 15, 43] with the assumptions $\tilde{m}/n_c \ll 1$ and $n_c \approx n$ for $T \to 0$. Then on the basis of Eq.(2), we obtain for the single-particle anomalous correlation function

$$\tilde{m}(\vec{r},0) = n \exp\left[-\frac{1}{2n}\int d\vec{k}\, \frac{\varepsilon_k}{E_k} \coth(\varepsilon_k/2T)\cos^2\left(\frac{\vec{k}.\vec{r}}{2}\right)\right], \quad (18)$$

At low temperatures ($T \ll \mu$) the main contribution to the integral of Eq.(18) comes from the region of small momentum then the single-particle anomalous correlation function undergoes a slow law decay at large distances:

$$\tilde{m}(r) = n\left(\frac{\xi}{r}\right)^{T/2T_d}, \quad (19)$$

where $T_d = \dfrac{2\pi\hbar^2 n}{m}$ is the temperature of quantum degeneracy.

We can infer from these results that the anomalous average does not exist at finite temperature. This is strictly confirmed by Eq.(19) where one finds that $\tilde{m}(r)$ vanishes for $r \to \infty$. Similarly to the situation in 2d at finite temperature regarding the normal correlator $\tilde{n}(\vec{r},\vec{r}') = <\Psi^+(\vec{r})\Psi(\vec{r}')> - \Phi^*(\vec{r})\Phi(\vec{r}')$, which tends to zero as $r \to \infty$, confirming that there is no true condensate, but one identifies instead the existence of a quasicondensate. However, this result also implies that there is no symmetry breaking, and consequently the anomalous average should not exist at any nonzero temperature. The butter of this result is that the anomalous density accompanies in a manner analogous the true condensate in a system of 2d homogeneous Bose gas.

## V. Superfluid fraction

Usually, Bose-Einstein condensation is accompanied by superfluidity. However, in a two-dimensional system at finite temperature, there is no BEC, but there still exists superfluidity. The relation between them depends on the Bogoliubov-type nature of the spectrum Eq.(7) [13]. Also, what is important is that our formalism provides a useful relation between the superfluid fraction and the dissipated heat which is equivalent to that obtained in Refs [13, 19].

$$f_s = \frac{n_s}{n} = 1 - \frac{\hbar^2}{2mnT}\int \frac{d^2k}{(2\pi)^2} k^2 \frac{e^{\varepsilon_k/T}}{\left(e^{\varepsilon_k/T}-1\right)^2} = 1 - \frac{Q_{d=2}}{T}, \quad (20)$$

where $n_s$ is the superfluid density.

It is very important to mention that the superfluid fraction $f_s$ will be a divergent quantity and thus the superfluid transition does not occur when the anomalous average is omitted in Eq.(20).

At low temperature and weak interaction, we get

$$f_s = 1 - \frac{3\varsigma(3)}{2\pi\hbar^2 mnc_s^4} T^3, \qquad (21)$$

where $\varsigma(3)$ is a Riemann zeta function and the sound velocity turns out to be given

$$c_s = c_s^{(0)} \sqrt{\left(1 - \frac{\tilde{n}}{n} + \frac{\tilde{m}}{n}\right)}, \qquad (22)$$

with $c_s^{(0)} = \sqrt{\mu_0 / m}$ is the zero order sound velocity.

Upon neglecting the normal and the anomalous fractions we recover straightforwardly the superfluid fraction obtained earlier by Popov [12] and by Fisher and Hohenberg [13].

## VI. Conclusion

We have studied in this paper the behavior of the anomalous density in two-dimensional homogeneous Bose gases. We find that this quantity has a finite value in the limit of weak interactions. We have discussed also the effects of the anomalous average on some thermodynamic quantities. As an example, we have given formulas for the chemical potential, ground-state energy, the depletion and superfluid fraction. The later does not occur if the anomalous density is neglected. In the ultra-dilute limit, the known results are reproduced. This feature makes our predictions in accordance with Monte-Carlo simulations and analytical calculations. Also, we have shown that our approach satisfies the HP theorem at zero temperature while it does not verify the Nepomnyashchy identity as it should be for any limited approximation. Moreover, the importance of the anomalous density compared to the normal one at low temperature has been also highlighted. In addition, by using the density phase fluctuation we found that the single-particle anomalous correlation function undergoes a slow law decay at large distances such a result implies that the anomalous average does not exist at finite temperature.

Finally, an interesting question to ask is whether some quantity can exist in this system which accompanies the quasicondensate density in a manner analogous to that in which the anomalous density accompanies a true condensate?

The goal of our next work is to use our approach to answer this important question. On the other hand we will try to extract something useful about superfluidity in 2d Bose system. The idea is to relate our predictions with appropriate numerical simulations for some realistic experiments in traps to study for example the vortex stability without rotating fluid.

**Acknowledgments**

We acknowledge Ludovic Pricoupenko, Gora Shlyapnikov, Jean Dalibard and Usama Al-Khawaja for many useful comments about this work. We are grateful to J. Andersen and V. Yukalov for helpful discussions. We are indebted to Yvan Castin for giving us the numerical data.